\documentclass{aastex}
\doublespace

\usepackage{epsfig}
\usepackage{subfigure}
\usepackage{verbatim}
\usepackage{enumerate}
\usepackage{color}
\usepackage{epstopdf}
\usepackage{courier}
\usepackage{natbib}
\usepackage{amsmath}

\begin{document}

\title{Projection effects in coronal dimmings and associated EUV wave event}
\author{K.~Dissauer\altaffilmark{1}, M.~Temmer\altaffilmark{1}, A.~M.~Veronig\altaffilmark{1,2}, K.~Vanninathan\altaffilmark{1} and J.~Magdaleni\'{c}\altaffilmark{3}}
\email{karin.dissauer@uni-graz.at}
\altaffiltext{1}{IGAM/Institute of Physics, University of Graz, Universit\"atsplatz 5/II, A-8010 Graz, Austria}
\altaffiltext{2}{Kanzelh\"ohe Observatory/Institute of Physics, University of Graz, A-9521 Treffen, Austria}
\altaffiltext{3}{Solar-Terrestrial Center of Excellence-SIDC, Royal Observatory of Belgium, Av. Circulaire 3, B-1180 Brussels, Belgium}
\begin{abstract}
We investigate the high-speed ($v >$ 1000~km~s$^{-1}$) extreme-ultraviolet (EUV) wave associated with an X1.2 flare and coronal mass ejection (CME) from NOAA active region 11283 on 2011 September 6 (SOL2011-09-06T22:12). This EUV wave features peculiar on-disk signatures, in particular we observe an intermittent ``disappearance'' of the front for 120~s in SDO/AIA 171, 193, 211~{\AA} data, whereas the 335~\AA~filter, sensitive to hotter plasmas (T$\sim$ 2.5~MK), shows a continuous evolution of the wave front. The eruption was also accompanied by localized coronal dimming regions. We exploit the multi-point quadrature position of SDO and STEREO-A, to make a thorough analysis of the EUV wave evolution, with respect to its kinematics and amplitude evolution and reconstruct the SDO line-of-sight (LOS) direction of the identified coronal dimming regions in STEREO-A. We show that the observed intensities of the dimming regions in SDO/AIA depend on the structures that are lying along their LOS and are the combination of their individual intensities, e.g. the expanding CME body, the enhanced EUV wave and CME front.
In this context, we conclude that the intermittent disappearance of the EUV wave in the AIA 171, 193, 211~\AA~filters, which are channels sensitive to plasma with temperatures below $\sim$ 2~MK is also caused by such LOS integration effects. These observations clearly demonstrate that single-view image data provide us with limited insight to correctly interpret coronal features.
\end{abstract}


\section{Introduction}
Coronal mass ejections (CMEs) are large-scale coronal magnetic field structures expelled into the heliosphere. They are the main causes of space weather disturbances since they may drive interplanetary shocks and produce geomagnetic storms. Accompanied with CMEs we observe activities low in the solar atmosphere including filament eruptions, flares, large-scale coronal EUV waves, and coronal dimmings.

EUV waves are large-scale disturbances propagating through the solar atmosphere, observed as moving fronts of increased coronal EUV emission.
They propagate at typical speeds of 200--400\,km\,s$^{-1}$ \citep[][]{Klassen:2000, Thompson:2009, Muhr:2014} but on rare occasions, EUV wave events with speeds in excess of 1000\,km\,s$^{-1}$ have been reported \citep[][]{Nitta:2013}.
Several models have been proposed to explain the nature of EUV waves, falling into three categories: wave, non-wave and hybrid models. 
Most wave models interpret the propagating signatures as fast-mode magnetosonic waves \citep[][]{Thompson:1999, Wang:2000, Wu:2001, Warmuth:2001, Ofman:2002}, while in the non-wave models they are explained as disturbances due to sucessive restructuring of magnetic field lines during the erupting CME \citep[][]{Delannee:1999, Chen:2002,Attrill:2007}. Hybrid models include both interpretations resulting in a bimodal picture, where a fast-mode wave travels ahead of a slower inner component due to magnetic field reconfiguration caused by the erupting CME \citep[][]{Zhukov:2004}.
A detailed discussion on the various theories and the evidence for and against them may be found in recent reviews \citep[e.g.][]{Warmuth:2010, Gallagher:2011, Zhukov:2011, Patsourakos:2012, Liu:2014, Warmuth:2015}. 

The relationship between CMEs and EUV waves was investigated by several statistical studies \citep[][]{Biesecker:2002, Cliver:2005, Nitta:2013, Nitta:2014, Muhr:2014}. It was found that fast and wide CMEs are in general accompanied by well-observed EUV waves that are associated with shocks, i.e. related to type II radio bursts \citep[][]{Cliver:2005}.
The unprecedented multi vantage-point observations from the \textit{Solar TErrestrial RElations Observatory} \citep[STEREO,][]{Kaiser:2008} and the high-cadence imagery by the \textit{Atmospheric Imaging Assembly} \citep[AIA,][]{Lemen:2012} on-board the \textit{Solar Dynamics Observatory} \citep[SDO,][]{Pesnell:2012} improved our understanding of how the EUV wave formation and kinematics are related to the CME dynamics. Initially the EUV wave front is found to be closely attached to the laterally expanding CME flanks. 
After an impulsive driving phase, the CME flanks decelerate, and the EUV wave subsequently becomes freely propagating with a velocity close to the fast magnetosonic speed in the quiet corona \citep[][]{Warmuth:2004, Long:2008, Veronig:2008, Patsourakos:2009, Patsourakos:2009a, Kienreich:2009, Veronig:2010, Olmedo:2012, Cheng:2012}. Quadrature observations of the two STEREO spacecraft unambiguously revealed that the EUV wave is propagating ahead of and is driven by the laterally expanding CME flanks \citep[][]{Patsourakos:2009a, Kienreich:2009, Ma:2011}. These findings were also confirmed by MHD simulations of CME eruptions and associated EUV waves \citep[][]{Pomoell:2008, Cohen:2009, Downs:2011, Downs:2012}.

Coronal dimmings are regions of reduced emission in the low corona observed in EUV \citep[][]{Thompson:1998,Thompson:2000} and soft X-rays (SXR; \citealt{Hudson:1996,Sterling:1997}). The appearance of these dimming regions is in general interpreted as density depletion caused by the evacuation of plasma during the CME expansion \citep[][]{Hudson:1996,Thompson:1998, Harrison:2000}. This interpretation is supported by the simulatenous and co-spatial observations of coronal dimmings in different wavelengths \citep[e.g.][]{Zarro:1999}, spectroscopic observations showing plasma outflows in dimming regions \citep[][]{Harra:2001,Attrill:2010,Tian:2012} and studies on dimming/CME mass relations \citep[][]{Sterling:1997, Wang:2002, Harrison:2003, Zhukov:2004, Mandrini:2007, Aschwanden:2009, Miklenic:2011}. In the literature, two different types of dimmings are defined, core (or twin) dimmings and secondary (or remote) dimmings, respectively. Core dimmings are stationary, localized regions that are often present on both sides of an erupting configuration, in opposite magnetic polarity regions. In relatively simple cases they are interpreted to mark the footpoints of the ejected fluxrope \citep[][]{Sterling:1997, Thompson:2000, Webb:2000, Mandrini:2005,Mandrini:2007, Temmer:2011}. The more shallow secondary dimmings are diffuse and can extend to significant distances from the source region. They are often observed to follow behind a propagating EUV wavefront
\citep[][]{Delannee:1999, Wills-Davey:1999, Thompson:2000, Attrill:2007, Mandrini:2007,Muhr:2011}. Therefore, they could be rarefaction regions that develop behind a compressive wave \citep[e.g.][]{Wu:2001, Muhr:2011, Downs:2012} or formed due to the plasma evacuation behind the flux rope and overlying fields that are erupting.

In this paper, we analyze the flare-CME event and associated large-scale EUV wave that occurred on \mbox{2011 September 6}. Several aspects of this event have been studied before. For instance, \cite{Nitta:2013} investigated the EUV wave kinematics using high-cadence observations from SDO/AIA, and determined this event to be a high-speed EUV wave with a velocity of $v\approx$\,1250\,km\,s$^{-1}$. Only 6 out of 171 events reported by \cite{Nitta:2013} were identified in this high-speed regime ($v>$1200\,km\,s$^{-1}$). The globally propagating EUV wave caused oscillations of filaments and the launch of a jet \citep{Shen:2014}. These results indicate that EUV waves may serve as agents for linking successive solar activities. \cite{Jiang:2013} focused on the triggering mechanism of the erupting event using MHD simulations initialized by a non-linear force-free field (NLFFF) extrapolation. \cite{Janvier:2016} studied the morphology and time evolution of photosperic traces of the current density and flare ribbons and compared it with the topological features found by NLFFF modeling. Both, \cite{Jiang:2013} and \cite{Janvier:2016} identified a spine-fan configuration of the overlying field lines, due to the presence of a parasitic positive polarity, embedding an elongated flux rope. The energy accumulation of the flare and the accompanied CME was studied by \cite{Feng:2013}. They found that the calculated free magnetic energy is able to power the flare and the CME, and that both phenomena may consume a similar amount of free energy. \cite{Romano:2015}  studied the evolution of the source active region (NOAA 11283) in relation to its recurrent flaring and CME activity. They found that before the occurence of the X2.1 flare-CME event under study here, the shearing motions seem to inject a larger fraction of energy into the corona than the emergence of the magnetic field.

In this paper, we investigate this extraordinary EUV wave and associated flare-CME event in detail and fully exploit the quadrature view from SDO and STEREO. 

\section{Observations and data analysis}\label{data}
\subsection{Event overview}
\begin{figure*}
     \centering
     \includegraphics[width=1.0\textwidth]{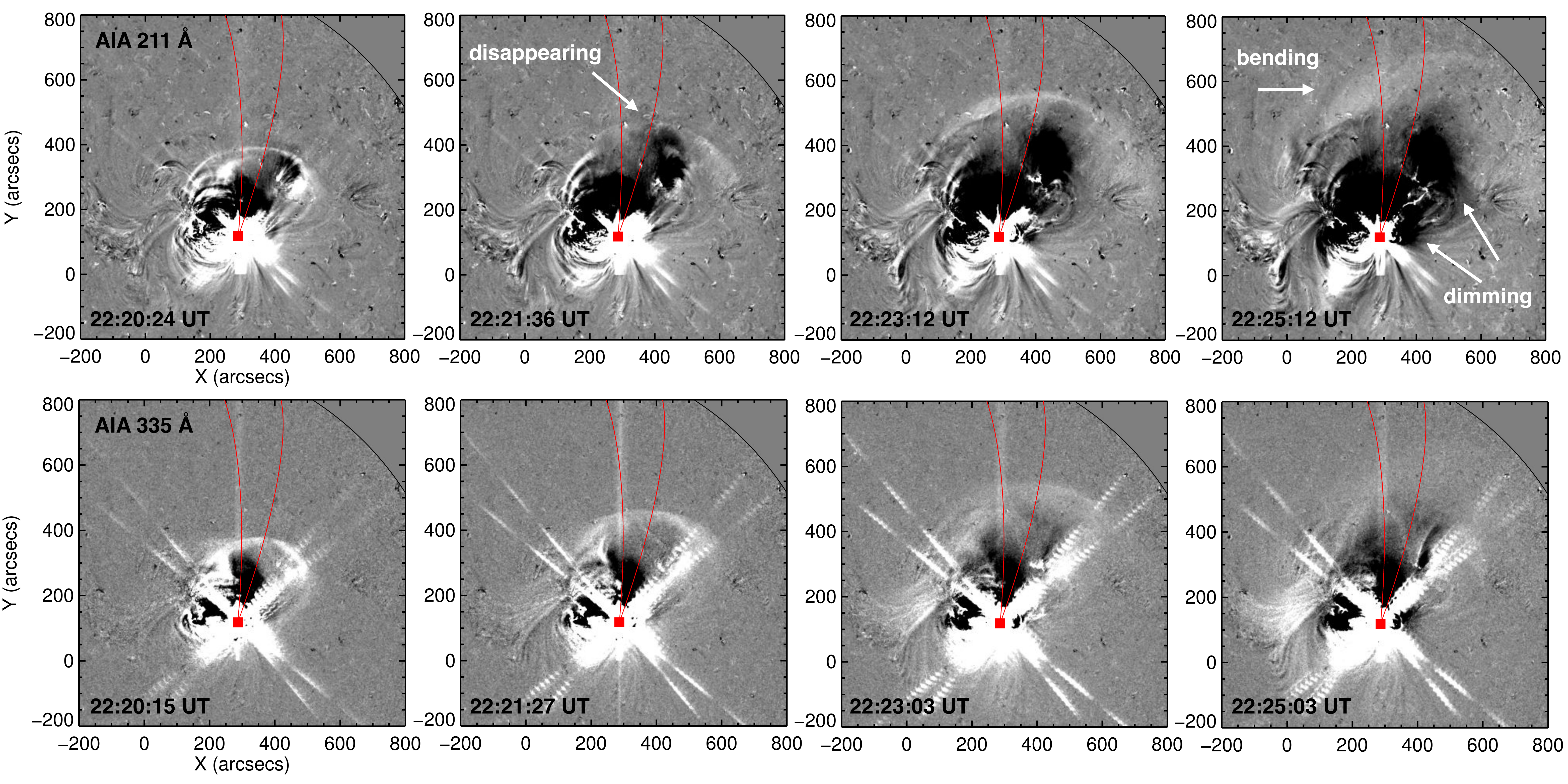}
\caption{Sequence of SDO/AIA base-difference images showing the evolution of the EUV wave followed by extended coronal dimming regions (top panels: 211~\AA; bottom panels: 335~\AA; see also the accompanied animation~no.~1). White arrows mark peculiar features such as the ``disappearance'' of a segment of the wave front in the 211~\AA~channel, as well as the localized dimming regions. Red lines indicate the sector of interest along which perturbation profiles are calculated.}
	\label{fig:event}
\end{figure*}

On 2011 September 6, a globally propagating high-speed EUV wave occurred in association with an X2.1 flare-CME event in NOAA active region 11283, at heliographic position N14$^{\circ}$W18$^{\circ}$ (cf. Fig.~\ref{fig:event}). The GOES and RHESSI X-ray lightcurves show the short impulsive phase of the flare starting at 22:16~UT. The GOES emission peaked at $\sim$22:21~UT (cf.\ top panel of Fig.~\ref{fig:sdokinematics}). RHESSI shows two intense hard X-ray (HXR) bursts up to 300~keV between 22:18 and 22:20~UT.

The EUV wave is most pronounced toward the North and associated with a type II burst, indicating its shock nature. However, no signature of an associated Moreton wave \citep{Moreton:1960} was observed in H$\alpha$ data of the Global Oscillation Network Group (GONG) network. 
The associated CME was observed with the coronagraphs of SOHO/LASCO as well as STEREO-A(head) (ST-A). From Earth-view it was first detected as a slow halo CME, in the SOHO/LASCO-C2 coronagraph at around 23:06~UT, with an average speed of $v$\,=\,580~km~s$^{-1}$ (CDAW catalogue). The speed as derived from ST-A, observing the CME almost in its plane of sky, is much higher $v$\,=\,990~km~s$^{-1}$ indicative of strong projection effects (cf.\ Sect. \ref{cmekin}). The eruption was also accompanied by the appearance of coronal dimming regions (cf.\ Fig.~\ref{fig:event} and movie no.~1).

\subsection{Data}

This event was observed in quadrature by the AIA instrument aboard SDO and the \textit{Sun Earth Connection Coronal and Heliospheric Investigation} \citep[{\it SECCHI};][]{Howard:2008} aboard ST-A, separated by 103$^{\circ}$.
With the eruption site being located at a longitude of W18$^{\circ}$, ST-A views the event over the West limb. Therefore, it is reasonable to assume that there are only small projection effects for measurements above the limb in ST-A imagery. The eruption site is located behind the limb for STEREO-B(ehind) (ST-B), hence data from this vantage point are not included in our analysis.
To study the EUV wave, we use high-cadence (12~s) observations from four SDO/AIA EUV channels. 171~\AA, 193~\AA~and 211~\AA~ are most sensitive to temperatures around $\approx 0.6-2 \times 10^{6}$~K, representing the quiet solar corona as well as plasma of active regions, while the SDO/AIA 335~\AA~ bandpass (peak formation temperature $\approx 2.5 \times 10^{6}$~K) is sensitive to hotter plasma in the corona \citep[][]{Lemen:2012}. 

We only use AIA images where the exposure times were constant and not triggered by the automatic exposure control algorithm of the instrument, resulting in a cadence of 24~s.
In order to enhance the faint structures of EUV waves and coronal dimmings and to study the relative as well as absolute changes in the emission, we construct base-ratio and base-difference images. To this aim, each image is divided or subtracted  by a pre-event image (recorded at 22:05~UT). We also use frames of running-difference images, where we subtracted from each image a frame that was taken 5 minutes before.
To compare the kinematics of the EUV wave obtained from different vantage points and to study the early evolution of coronal dimmings, we combine ST-A/EUVI 195~\AA~observations (cadence of 5~min) with SDO/AIA 193~\AA~data. 
We derive the kinematical evolution of the CME combining imagery from the SECCHI instrument suite, EUVI, COR1, and COR2 from ST-A.
All data were prepared using standard Solarsoft IDL software (\texttt{aia\_prep.pro} and \texttt{secchi\_prep.pro}) and each data set was differentially rotated to a reference time of 22:05~UT using \texttt{drot\_map.pro}.

\subsection{Perturbation Profiles}
We study the EUV wave pulse quantitatively using the perturbation profile method \citep[e.g.][]{Warmuth:2004, Podladchikova:2005, Veronig:2010, Muhr:2011, Long:2011}. The method allows a detailed study of the temporal evolution of the propagating wave front and its changes in amplitude by generating perturbation profiles along a specific sector.  
We derive the perturbation profiles from base-ratio images by calculating the median of the relative intensity changes of all pixels that are within the selected sector of propagation and successive annuli of $1^{\circ}$ width around the origin of the wave. The median absolute deviations are used as 1$\sigma$ measure errors. 
To identify the characteristics of the EUV wave a Gaussian curve is fitted to the positive section of each profile.
MPFIT \citep[][]{Markwardt:2009} is used to calculate the parameters of the Gaussian function with error bars. The Gaussian function that is fitted to the data is of the form
\begin{equation}
I(x)=p_{0}+p_{3}\, \exp{\left( -\frac{\left(x-p_{1}\right)^{2}}{2 p_{2}^{2}}\right)} \; \vspace{0.5em},
\end{equation}
where, $p_{0}\ldots p_{3}$ are the best-fit values returned. We use the perturbation profiles to study the kinematics of the EUV wave pulse as well as the time evolution of its amplitude.

\subsection{Projection and LOS integration}
Since the solar corona is an optically thin medium, the measured EUV intensity (e.g. from instruments aboard SDO or STEREO) results from integration of all the emission along the LOS.
The simultaneous observations of two spacecraft (SC1, SC2) allow us to interpret the intensity of each image pixel of SC1 as the sum over the intensities of all image pixels of SC2 that lie along the projected LOS. We note that this is only a qualitative approach to identify which structures lie along the LOS of selected locations, since also the image data of SC2 itself results from projection and LOS integration.
For our purpose, we select regions of interest in SDO image data and overplot the projection of the three-dimensional LOS onto the solar surface as observed by ST-A using the IDL routine \texttt{scc\_measure.pro} \citep[][]{Thompson:2009b}. To transform positions of SDO measurements to ST-A, we adopt the World Coordinate System (WCS) routines in SSWIDL \citep[][]{Thompson:2010}.
We use this approach to investigate which structures, observed by ST-A on the solar limb, lie along the LOS of on-disk locations of interest of SDO, such as the position of the EUV wave front and the coronal dimmings regions.

\section{Results}\label{results}
\begin{figure*}
  \begin{minipage}{1.0\textwidth}
     \centering
     \includegraphics[width=1.0\textwidth]{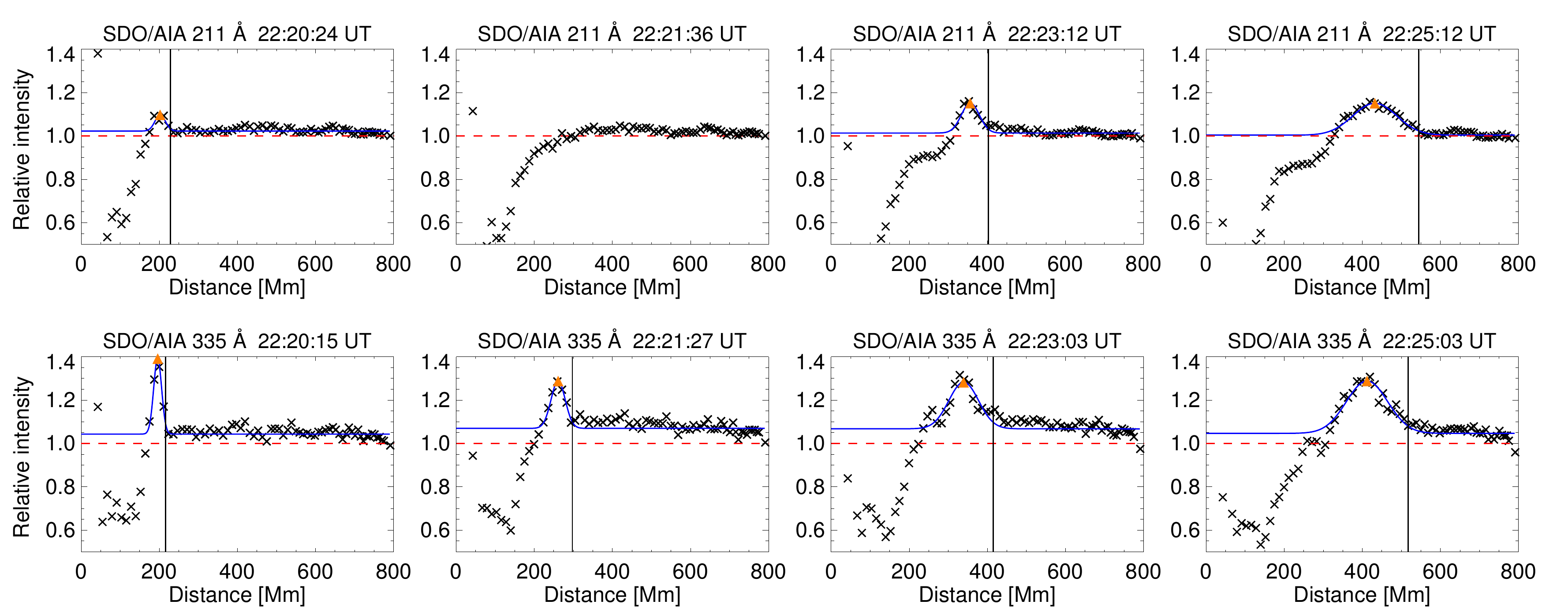}
 \end{minipage}
  \caption{Perturbation profiles derived from SDO/AIA 211~\AA~(top panels) and 335~\AA~(bottom panels) base-ratio images along the $15^{\circ}$ wide sector indicated in Fig.~\ref{fig:event}. The blue curve outlines the Gaussian fit to the data. The maximum of the Gaussian fit (relative amplitude) is marked by the orange triangle (wave peak), while the vertical line corresponds to the leading edge of the front determined from the fit. The ``disappearance'' of the wave front can be seen in the 211~\AA~channel (second panel), while the wave pulse evolves continuously in the 335~\AA~channel. An animation of this figure is available.}
  \label{fig:profiles}
\end{figure*}
\begin{figure}
  \begin{minipage}{1.0\textwidth}
\centering
     \includegraphics[width=0.6\textwidth]{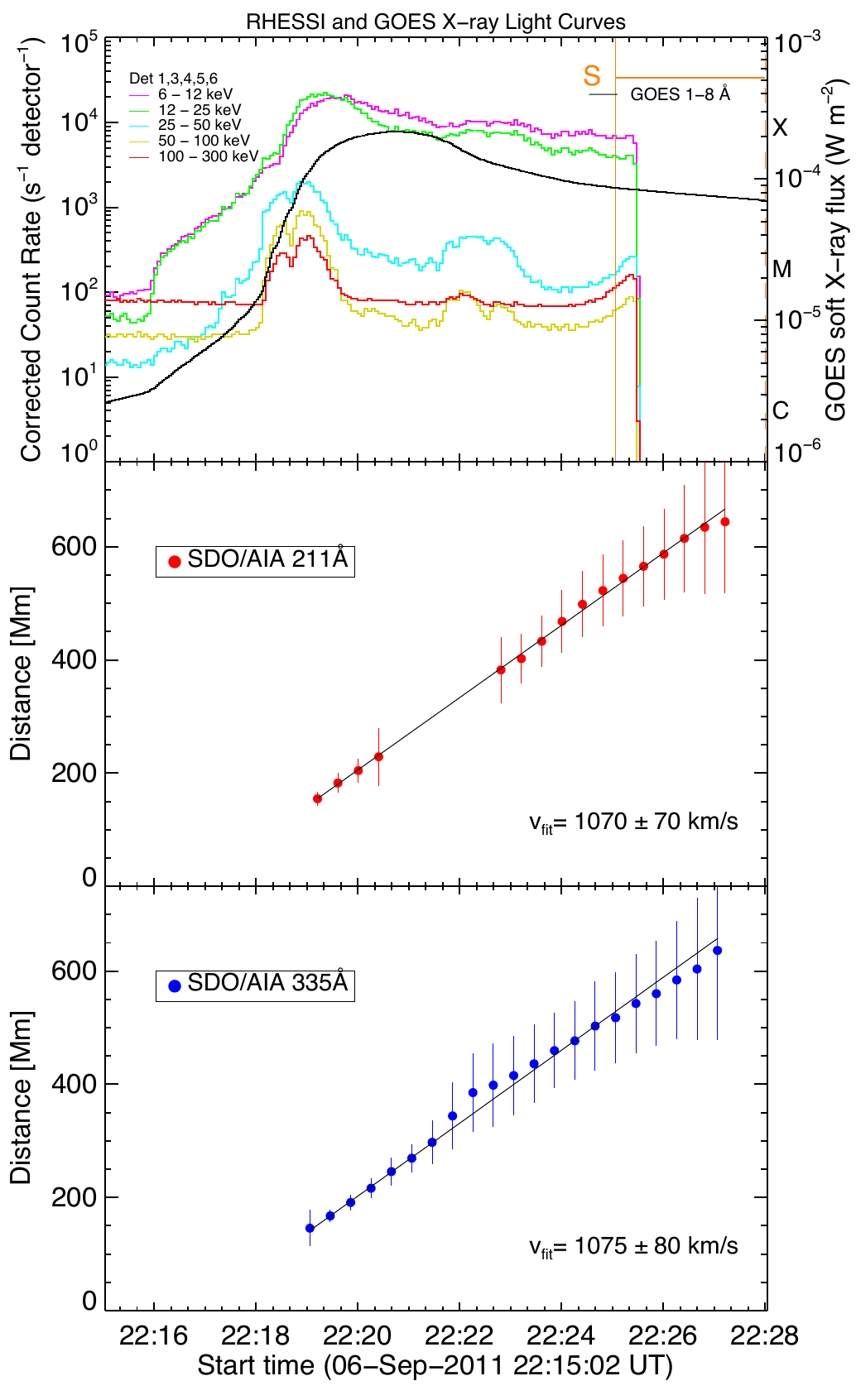}

  \caption{Top: RHESSI HXR count rates in five energy bands from 6~keV to 300~keV and GOES 1.0-8.0~\AA~SXR flux (black solid line) during the X2.1 flare on 6 September 2011. Middle and bottom: kinematics of the EUV wave front obtained from perturbation profiles calculated from SDO/AIA 211~\AA~and 335~\AA~ data. Since no proper identification of the wave front was possible during the times of disappearance for the 211~\AA~channel, measurements during these times were excluded.}
 \label{fig:sdokinematics}
\end{minipage}
  \end{figure}

Figure~\ref{fig:event} shows the evolution of the EUV wave in SDO/AIA 211~\AA~and 335~\AA~base-difference images (see also the accompanying animation~no.~1). The EUV wave occurs as a sharp front, with a first appearance at 22:19~UT and can be followed up to about 22:27~UT. As can be seen in the 211~\AA~images, the segment of the wave front propagating towards the North ``disappears'' at 22:20:48~UT, and then reappears at 22:22:48~UT. The same behaviour is observed in the 193~\AA~and 171~\AA~channels (not shown). In contrast, the entire wave front remains clearly visible in the 335~\AA~channel. Coronal dimming regions are observed behind the EUV wave front. The shape and extension of the dimming regions appears differently in various SDO/AIA channels (e.g.\ 211~\AA~ vs. 335~\AA; Fig.~\ref{fig:event}).

\subsection{EUV wave}
To derive the kinematical evolution of the EUV wave we calculate perturbation profiles for the direction in which we observe the disappearance of the wave front over a sector of $15^{\circ}$ angular width (indicated with red lines in Fig.~\ref{fig:event}).
Figure~\ref{fig:profiles} shows snapshots of the time evolution of these profiles for the 211~\AA~and 335~\AA~channels (see also animation no.~2). To determine the velocity of the EUV wave, we define the leading front of the wave \mbox{$x_{\text{lead}}$} to be a function of the parameters of the Gaussian fit, namely
\begin{equation}
x_{\text{lead}}=p_{1}+2p_{2}\;.
\end{equation}
Accordingly, the $1\sigma$ errors in $x_{\text{lead}}$ are
\begin{equation}
\Delta x_{\text{lead}}=\pm \left(\Delta p_{1} + 2\Delta p_{2} \right) \;.
\end{equation}
The resulting wave kinematics for SDO/AIA 211~\AA~ and 335~\AA~ are plotted in the middle and bottom panels of Fig.~\ref{fig:sdokinematics}, respectively. The position of the leading edge of the wave front determined by this procedure (indicated by the vertical lines in Fig.~\ref{fig:profiles}) for each perturbation profile is plotted together with the $1 \sigma$ error against time. The velocities obtained from the linear fit to the time-distance data are \mbox{$v_{211}=1070\,\pm\,70$~ km~s$^{-1}$} for 211~\AA~and \mbox{$v_{335}=1075\,\pm\,80~$km~s$^{-1}$} for 335~\AA.

\begin{figure*}
 \begin{minipage}{1.0\columnwidth}
     \centering
     \includegraphics[width=0.9\columnwidth]{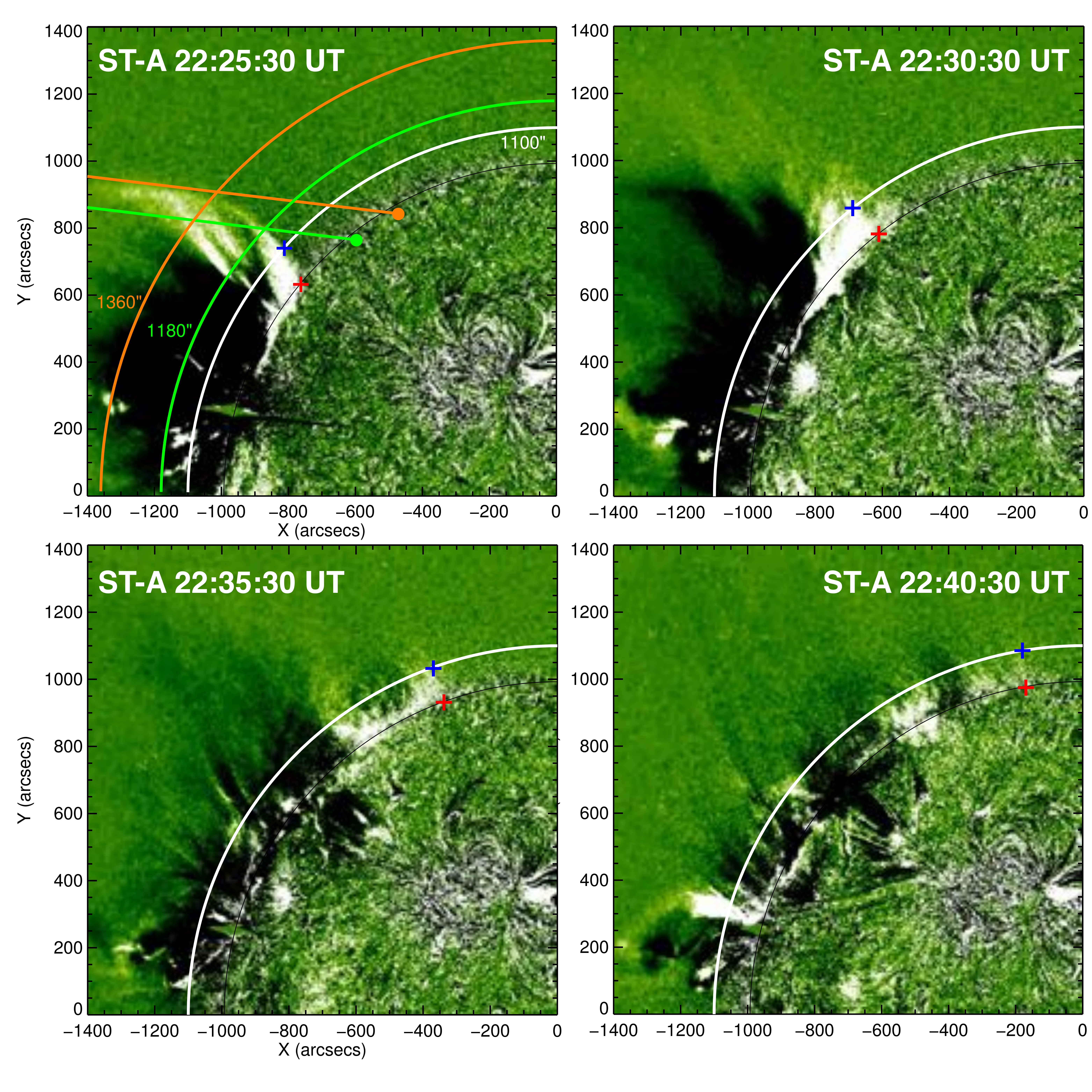}
 \end{minipage}
 \caption{Running-difference images showing the time evolution of the EUV wave and the expanding CME in ST-A/EUVI 195~\AA. Red crosses indicate the position of the wave front as measured on the solar limb. Blue crosses represent the distance of the wave front along a circle with radius 1100$\arcsec$ (i.e., a height of 130$\arcsec$ above the solar limb). The green and orange dots in the upper left panel mark the transformed position of the peak and the leading front of the EUV wave identified in the profiles from AIA 193~\AA~data at 22:25:31 UT. 
The overplotted lines represent the projected LOS for those measurements. The equivalent colored circles above the solar limb indicate at which heights the overplotted LOS match with the outermost visible front in ST-A.}
\label{fig:stereo}
\end{figure*}
\begin{figure}
 \begin{minipage}{1.0\columnwidth}
     \centering
     \includegraphics[width=1.0\columnwidth]{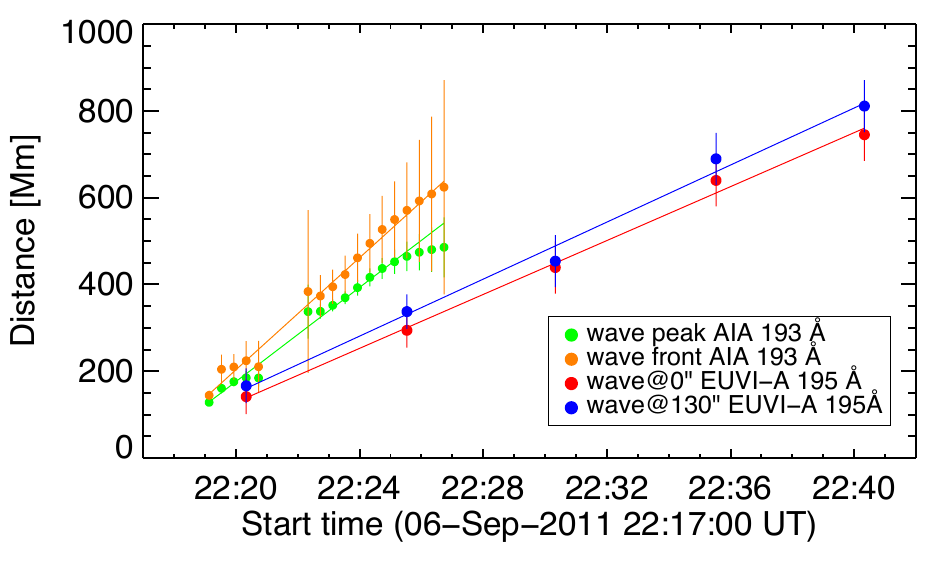}
 \end{minipage}
 \caption{Time evolution of the distance of the EUV wave front (from the wave center) obtained from ST-A/EUVI and SDO/AIA (red: EUVI-A 0$\arcsec$; blue: EUVI-A 130$\arcsec$ above the solar limb, green: peak of AIA; orange: leading front determined in AIA).}
\label{fig:sdostereokin}
\end{figure}

The observations from quadrature view enable us to study the wave signature simultaneously off-limb and on-disk. Fig.~\ref{fig:stereo} shows a sequence of running-difference images illustrating the evolution of the eruption in ST-A 195~\AA. The EUV wave front was tracked manually along the solar limb, indicated by the red crosses. Since the typical heights of an EUV waves have been established to be between 80-100~Mm \citep[][]{Patsourakos:2009,Kienreich:2009}, we also include measurements at a height of 90~Mm ($\sim$\,130$\arcsec$ above the solar limb, i.e. circle with radius 1100$\arcsec$, marked as blue crosses).  For SDO, we obtain the time evolution of the peak and the leading front of the EUV wave pulse from perturbation profiles for the 193~\AA~channel. Fig.~\ref{fig:stereo} also shows that initially the CME front cannot be distinguished from the EUV wave. Around 22:30~UT the lateral expansion of the CME stops and we identify the EUV wave propagating ahead of the CME.

Figure~\ref{fig:sdostereokin} shows the distance-time curves obtained from ST-A 195~\AA~(off-limb) and SDO 193~\AA~(on-disk). Different speeds are derived using mesurements obtained from different vantage points. From SDO, we derive a speed of $v_{\text{lead}}=1080\,\pm\,80$~km~s$^{-1}$ for the leading front and $v_{\text{peak}}=900\,\pm\,30~$km~s$^{-1}$ for the peak of the wave pulse. From ST-A view the resulting speeds are significantly lower, amounting to $v=520\,\pm\,40~$km~s$^{-1}$ for the leading front of the wave propagating along the solar limb, and $v=550\,\pm\,40~$km~s$^{-1}$ at a height of 90~Mm. These discrepancies are investigated in more detail. In Fig.~\ref{fig:stereo} we show the position of the EUV wave's peak and leading front as observed from SDO transformed to ST-A view (marked as green and orange dots, respectively). We clearly see, that both on-disk features do neither correspond to a wave signature on the solar limb, nor to a wave signature at a height of $\sim$ 90~Mm in ST-A. To estimate the actual heights, where the emission for those signatures come from, we plot also the projected LOS of the transformed SDO/AIA wave positions in ST-A (indicated by the green and orange line) in order to identify where the LOS matches with the outermost visible front in ST-A (indicated by the green and the orange circle). We obtain for the leading edge of the front a height of $\sim$\,260~Mm and for the peak of the wave pulse a height of $\sim$\,130 Mm above the solar limb. 

\begin{figure}
   \begin{minipage}{1.0\columnwidth}
\centering
\includegraphics[width=1.0\columnwidth]{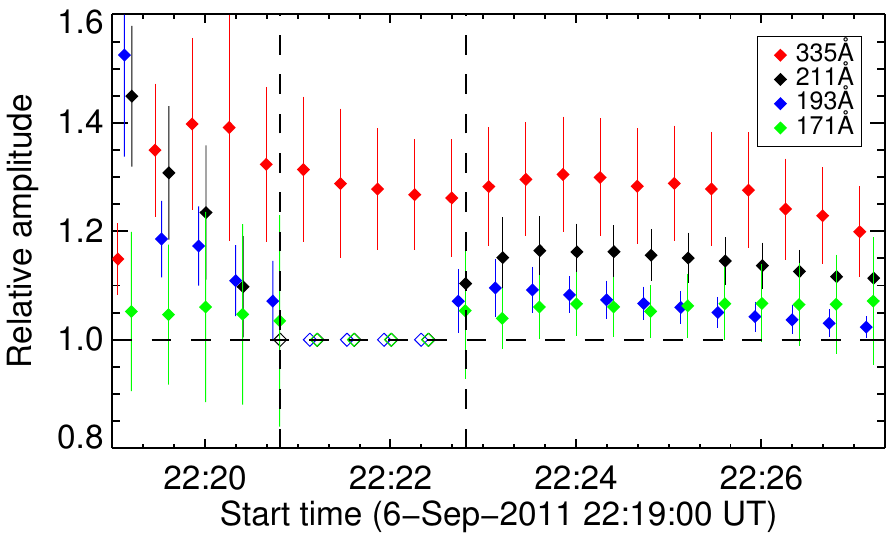}
  \end{minipage}
  \caption{Time evolution of the EUV wave amplitude determined from different AIA filters (see legend). The two vertical black dashed lines indicate the disappearance of the wave front at 22:20:48~UT as well as the reappearance at 22:22:48~UT identified in the 211~\AA~channel. No proper identification of the amplitude of the EUV wave was possible during these times for 211~\AA, 171~\AA~ and 193~\AA. Those measurements are marked with open symbols and represent the decrease of the relative intensity to the background level.}
  \label{fig:amplitude}
\end{figure}

Another interesting aspect of the EUV wave is the intermittent disappearance of parts of the front. Fig.~\ref{fig:amplitude} shows the amplitude of the Gaussian fit to each perturbation profile (cf. orange triangles in Fig.~\ref{fig:profiles}) plotted against time. The amplitude's time evolution as well as the perturbation profiles themselves (cf. Fig.~\ref{fig:profiles} and movie no.~2) clearly reveal that the EUV wave pulse ``disappears", i.e. the relative intensity decreases to the background level of 1.0, in the 211~\AA~filter at 22:20:48~UT, and reappears again at 22:22:48~UT. The same behaviour is observed for the 171~\AA~and the 193~\AA~channels. In contrast, the 335~\AA~profiles show a continuously propagating wave pulse, but the pulse amplitude reveals a decrease during the time of disappearance in the 211~\AA~filter (marked as black dashed vertical lines in Fig.~\ref{fig:amplitude}). After its ``reappearance'', the amplitude of the wave pulse increases in all four filters.

\subsection{CME and type-II radio burst}\label{cmekin}
\begin{figure}
  \begin{minipage}{1.0\columnwidth}
	\centering
	\includegraphics[width=1.0\columnwidth]{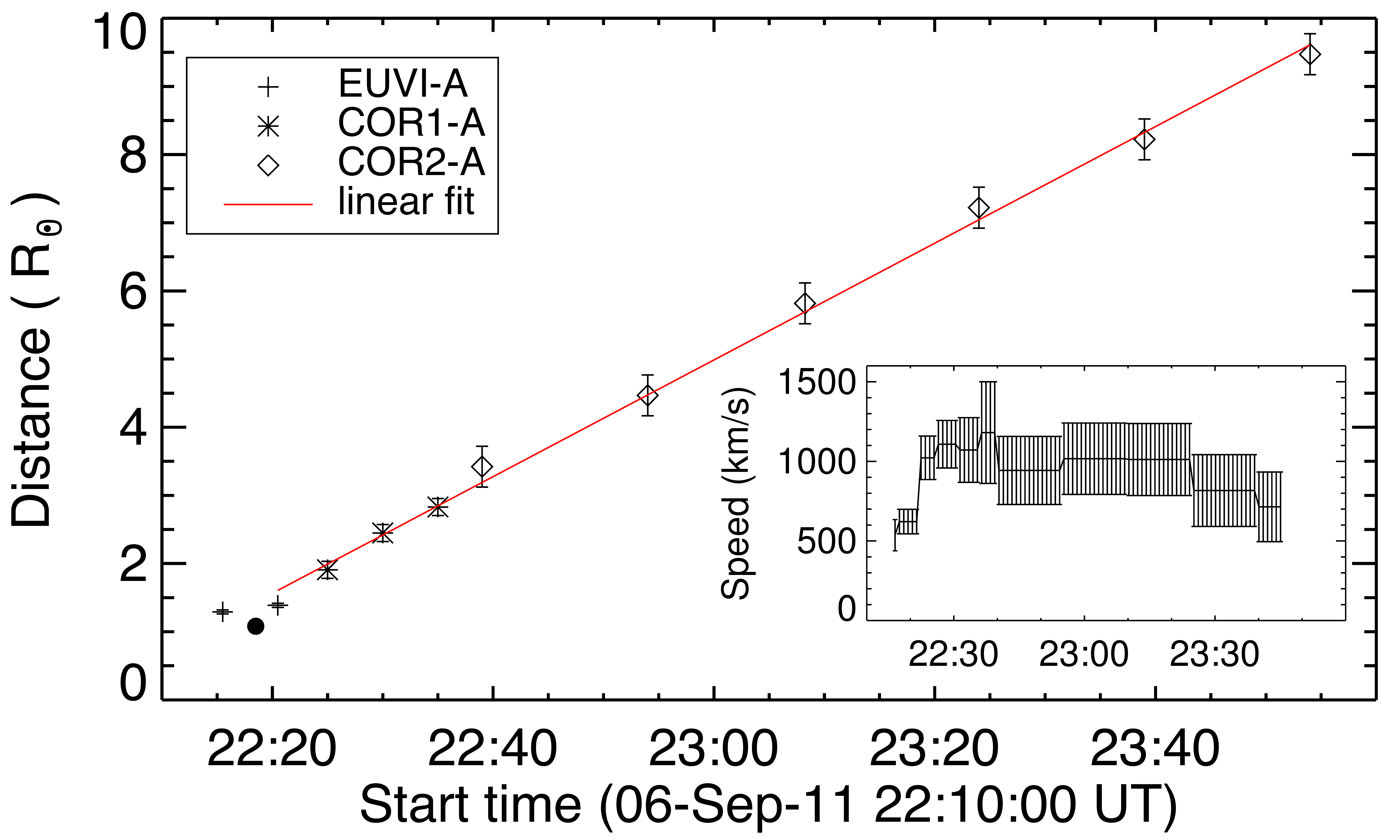} 
 \end{minipage}
 \caption{Distance-time plot of the CME front measured in ST-A/EUVI, COR1, and COR2 data. The red solid line indicates the linear fit to the measurements. The inset to the figure shows the speed-time profile derived by using the regularization method from \cite{Temmer:2010}. The black dot represents the reconstructed initiation height of the shock wave obtained from radio observations.}
 \label{cme}
 \end{figure}
 
Figure~\ref{cme} shows the CME kinematics obtained from ST-A/EUVI, COR1 and COR2 data by tracking the CME frontal part manually along its main propagation direction. Since from the ST-A view, the event occurred on the solar limb, these measurements of the CME propagation are basically uneffected by projection effects. We derive an average speed of $v=990\,\pm$ 50~km~s$^{-1}$, which is much higher than the value provided in the CDAW catalogue derived from SOHO/LASCO observations ($v=580$~km~s$^{-1}$), where the CME is observed as halo, i.e.\,propagating towards the observer. 

The event was also associated with a type II radio burst, observed by different ground-based instruments (Hiraiso Radiospectrograph, Culgoora Spectrograph, Green Bank Solar Radio Burst Spectrometer and Sagamore-Hill). The patchy and intermittent continuation of the coronal type II emission to the interplanetary space was observed by the Waves instrument on-board the WIND spacecraft (WIND/ Waves; \citealt{Bougeret:1995}) and the two Waves instruments on-board STEREO (STEREO/WAVES; \citealt{Kaiser:2005, Kaiser:2008, Bougeret:2008}). 
The type II radio burst, signature of a coronal shock wave, started concurrently with the EUV wave, at about 22:19 UT. 
The high starting frequency of the complex and structured type II burst (340~MHz for the fundamental emission band) indicates that the shock was formed low in the solar corona.

From the frequency drift rate and the generally used five-fold Saito coronal electron density model \citep[][]{Saito:1970} we obtain a shock wave speed of \mbox{$v=1020\,\pm$ 30~km~s$^{-1}$} and a starting height of \mbox{$h=1.08 \pm 0.02$~R$_{\odot}$, i.e.$\,\sim$\,60~Mm} above the solar surface (marked as black dot in Fig.~\ref{cme}). The study of \cite{Cho:2013} showed comparable results for the starting height of the high frequency type II radio burst. We note that the speed of the coronal shock is comparable to the speed of the EUV wave signatures as measured from SDO  and the CME speed obtained from ST-A observations (1070 and 990~km~s$^{-1}$, respectively).

\subsection{Coronal dimmings} 
To identify the coronal dimming regions, we apply a thresholding technique that uses for each pixel an upper limit on the absolute and relative decrease of the intensity with respect to the pre-event image. A pixel is flagged as dimming pixel if its intensity decreased by at least 20\% compared to its pre-event value and decreased by a certain absolute value A (in DN) from its pre-event value. Within this set of identified dimming pixels, we determine the 5\% ``darkest" pixels, i.e. the 5\% pixels that reveal the largest absolute change of their intensity. 
\begin{figure}
  \begin{minipage}{1.0\columnwidth}
     \centering
     \includegraphics[width=1.0\columnwidth]{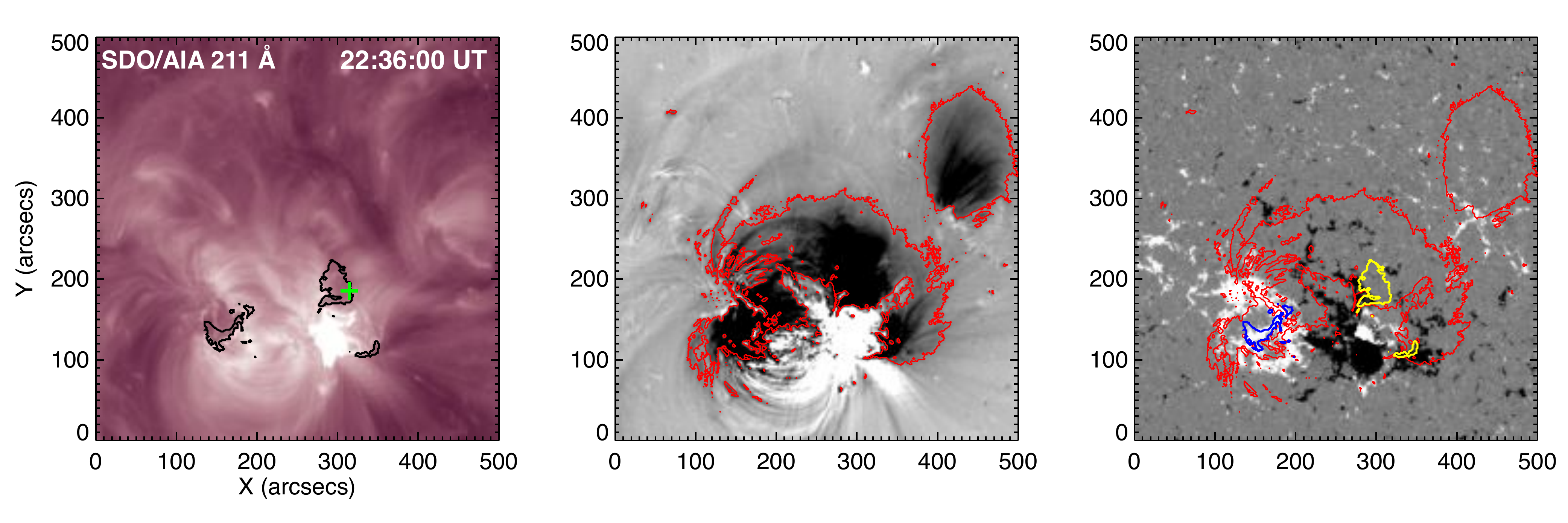}
 \end{minipage}
  \caption{Left to right: SDO/AIA 211~\AA~ direct image, corresponding base-difference image and SDO/HMI LOS magnetogram. The red contours indicate the overall coronal dimming regions identfied, while black contours mark regions of potential core dimmings. The core dimming regions (blue and yellow in the right panel indicating opposite magnetic polarities) are located in regions of opposite magnetic polarity and lie close to the eruption site. The location marked with the green cross (left panel) is reconstructed in ST-A (cf. Fig.~\ref{dimming_los}).}
  \label{dimming}
\end{figure}
Figure~\ref{dimming} illustrates the application of this method and shows a direct image in SDO/AIA 211~\AA, the corresponding base-difference image and the SDO/HMI LOS magnetogram at a time where the dimming region is still evolving. 
The red contours indicate the overall identified dimming regions. The black (left panel), blue and yellow contours (right panel) mark the 5\% darkest pixels. We interpret these regions as ``potential" core dimming regions, since it is assumed that dense plasma, previously confined by the flux rope, is evacuated there and the selected pixels show the maximum decrease in intensity of all identified dimming regions. 

As a next step, we exploit the quadrature configuration of SDO and ST-A to study the three-dimensional extension and position of the identified coronal dimming regions. To this aim, we select regions-of-interest (ROIs) in SDO/AIA observations and perform a coordinate transformation of these ROIs to ST-A/EUVI. 

\begin{figure}
  \begin{minipage}{1.0\columnwidth}
     \centering
     \includegraphics[width=1.0\columnwidth]{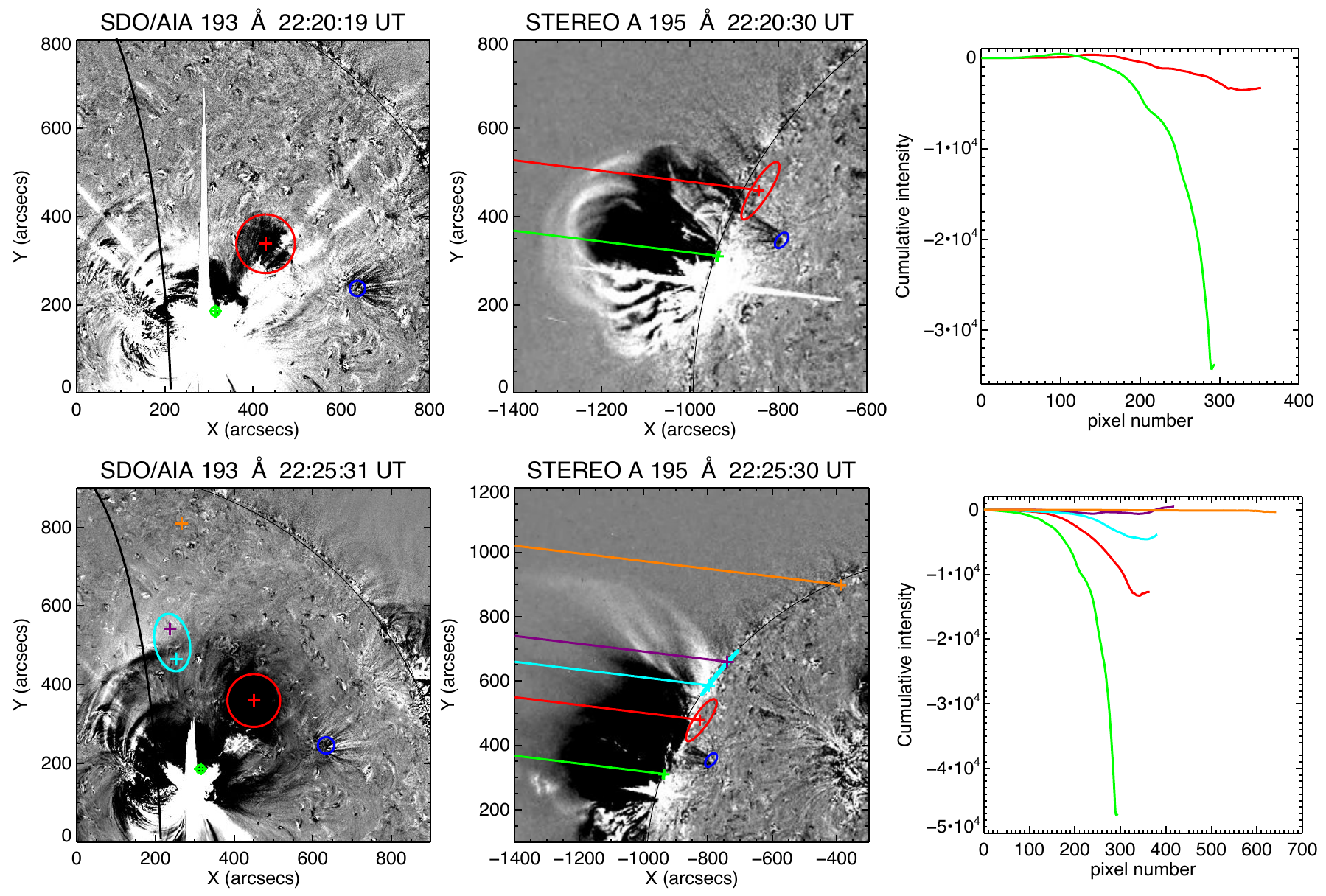}
 \end{minipage}
  \caption{Left panels: Base-difference images of the coronal dimming regions observed in the SDO/AIA 193~\AA~filter for two different time steps. The green circle marks a small subfield in a potential core dimming region (cf.~Fig.~\ref{dimming}), while the red one marks a region that appears to be a core dimming but turns out to result from a dimmed region higher up in the corona. The cyan oval marks a ROI that includes both enhanced and reduced intensity. The blue circle and the orange cross are used as a reference. Middle panels: corresponding base-difference images from ST-A/EUVI view, including the transformed on-disk signature of the ROIs selected from SDO/AIA. Right panels: ST-A 195~\AA~intensity cumulated along the colored lines in the middle panels (from left to right) indicating the SDO viewing direction.}
  \label{dimming_los}
\end{figure}

Figure~\ref{dimming_los} shows SDO/AIA 193~\AA~base-difference images at two different timesteps. We study areas within a potential core dimming region (green circle), within the overall dimming region (red circle) and a ROI that includes both enhanced and reduced intensity (cyan oval). To check for the correctness of the coordinate transformation, we mark as reference point an active region (blue circle). 
To investigate the origin of the contributions for the different dimming regions, we overplot the projected LOS for selected points inside each ROI in SDO/AIA (marked as crosses in Fig.~\ref{dimming_los}) onto the corresponding image of ST-A/EUVI. This is indicated by lines in the corresponding colors. The sum of all pixel intensities along each LOS is expected to correspond to the observed intensity in the marked locations in SDO/AIA. Indeed, this is qualitatively obtained from the right panels of Fig.~\ref{dimming_los}, where we plot the cumulative sum of all pixel intensities along the LOS for each selected location.
For instance, the position marked with the purple cross corresponds to a region of enhanced intensity in SDO/AIA; the ST-A view shows that the LOS crosses the CME flanks and EUV wave front.
The corresponding curve for the cumulative sum over the LOS pixels shows qualitatively the same result, i.e. an enhanced intensity compared to the background. The location marked with the green cross corresponds to a core dimming region and therefore to a location of much reduced intensity. As can be seen from its cumulative curve, only reduced intensity pixels (dimming pixels) lie along its LOS. 
The reconstruction of the LOS for the red cross reveals that the dimming contributions result from regions higher up in the corona (c.f. upper panels in Fig.~\ref{dimming_los}). The resulting emission is much higher than the reconstructed intensity of the core dimming region (green cross), indicating that in this dimming region lesser dense plasma is evacuated.
The cyan cross marks an even more shallow dimming region, which shows a smaller number of dimming pixels lying along the LOS as well as a contribution of pixels with enhanced intensities with respect to the pre-event state. The orange cross is used as a reference and as expected the reconstructed intensity equals to zero.
The right panels of Fig.~\ref{dimming_los} clearly show that the highest decrease in the emission reconstructed from ST-A comes from the core dimming region, while the other ROIs (c.f. red and cyan locations and the corresponding curves for the cumulative sum) show regions of secondary dimmings.
We note that this procedure is only approximately correct, as we can only make LOS cuts along the LOS-projected image of the approximate quadrature configuration, which by itself is also a result of LOS-integration of the emission observed. 

\section{Discussion and Conclusion}\label{discussion}

The EUV wave/CME/coronal dimming event on September 6, 2011 shows several peculiar features. Only by complementing the results of SDO observations with ST-A, we were able to disentangle real signatures from signatures arising from LOS integration and projection effects.
Furthermore, we were able to find new aspects for the formation of associated coronal dimming regions.

\begin{figure}
 \begin{minipage}{1.0\columnwidth}
     \centering
	\includegraphics[width=0.5\textwidth]{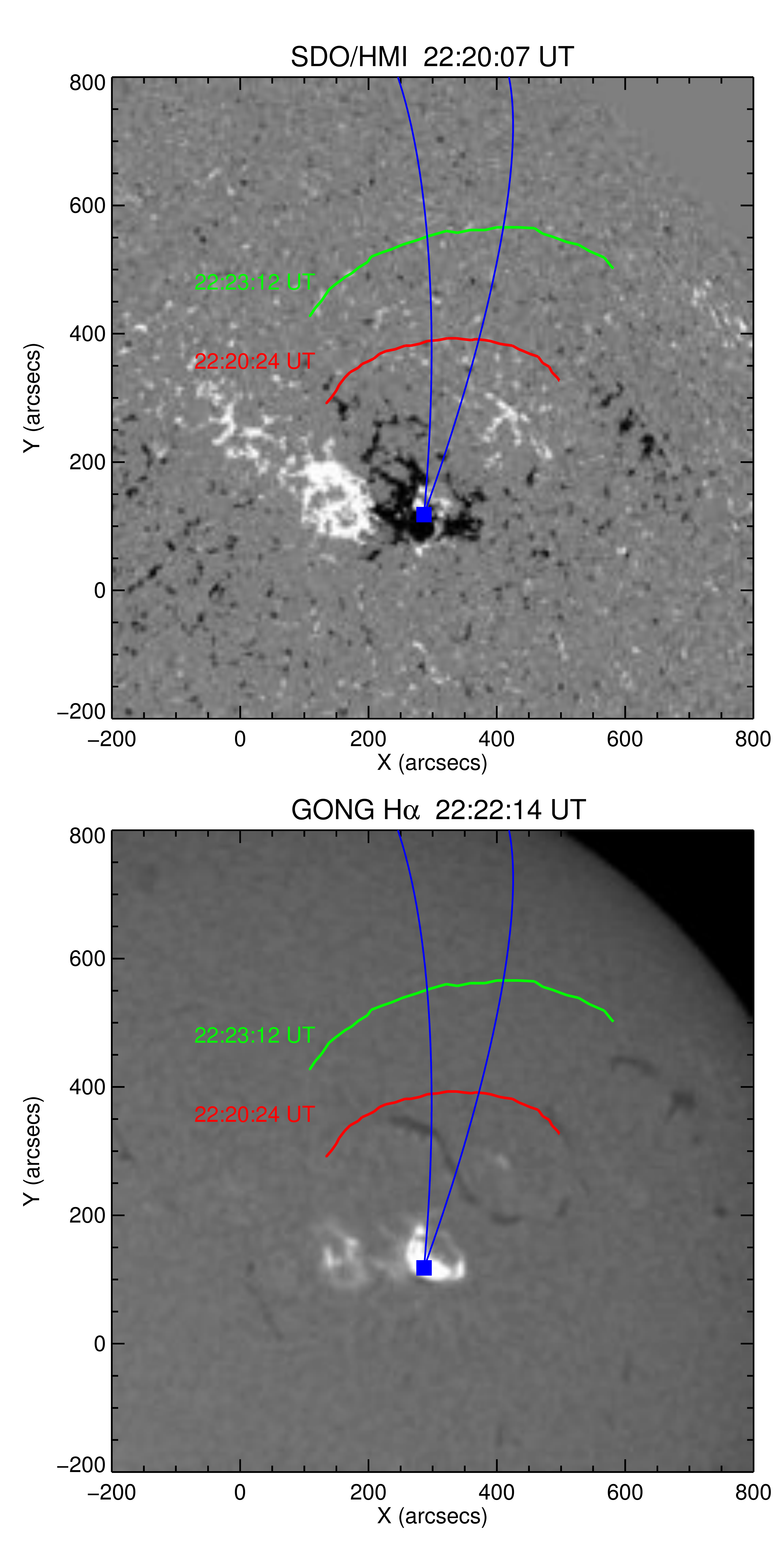}
\caption{Top: HMI magnetogram. Bottom: GONG H$\alpha$ filtergram showing the location of filaments. Overlaid on both panels is the measured sector (marked in blue), and the wave front determined at the different time steps that span the range of the wave ``disappearance'' (red and green line).}
	\label{hmi_mean_pos_neg}
\end{minipage}
\end{figure}

1) For the first time we identified the ``disappearance" of a segment of the EUV wave front followed by its reappearance and further propagation in the absence of any obvious magnetic structures. In the following we discuss three possible scenarios for this phenomenon: 

i) Heating at the wave front: \cite{Vanninathan:2015} studied the strong EUV wave associated with the X2.2 flare of 2011 February 15 and showed that the passing EUV front adiabatically compresses the ambient coronal plasma, resulting in an increase in density of about 6-9\% and an associated temperature increase of 5-6\%. Using the method described in \cite{Downs:2011}, we estimated for the event under study the intensity ratio for each AIA channel as a function of temperature, by assuming that the passing EUV wave causes $\sim$8\% increase in density compared to the pre-event corona (consistent with the values derived by \citealt{Veronig:2011, Schrijver:2011} and \citealt{Vanninathan:2015}). This leads to a $\sim$5\% increase in temperature. From the estimated intensity ratios we obtain that for a $\sim$5\% increase in temperature, the intensity in the 171~\AA~and 193~\AA~channels is reducing while that of 211~\AA~and 335~\AA~channels continue to increase. We repeated this test for different densities ranging from 6-12\% increase and obtained the same qualitative results in all cases. This is contradictory to the observed evolution of the amplitude of the wave pulse revealing a decrease in intensity for all four wavelength channels during its ``disappearance'' phase (cf.~Fig.~\ref{fig:amplitude}). This result is inconsistent with the disappearance being caused by heating at the wave front. 

ii) Wave propagation through an inhomogenous medium: 
EUV waves are known to be affected by inhomogeneities in the corona (in terms of Alfv\'en velocity), such as active regions, coronal holes or filaments. Encountering a region of high Alfv\'en speed would alter the amplitude of the wave pulse. 
To test this hypothesis, we checked for magnetic obstacles in SDO/HMI LOS magnetograms along the propagation path, which would correspond to regions of enhanced Alfv{\'e}n speed. Fig.~\ref{hmi_mean_pos_neg} shows LOS magnetic field and H$\alpha$ data together with the distance range, where the wave front ``disappears" (marked by the red and green fronts). No magnetic flux enhancements or filaments could be identified along the sector of interest. 
However, a filament is present along the propagation direction of the EUV wave but not at the locations where we observe the wave front disappearance. We note that the wave front passage causes the filament to oscillate, but it does not erupt \citep[][]{Shen:2014}. 
These observations suggest that the propagation through an inhomogenous medium is not responsible for the disappearance of the wave front.

iii) LOS effects: The simultaneous decrease in amplitude for all channels indicates that the decrease in emission results primarily from changes in the density and not changes in temperature. The 171~\AA, 193~\AA, and 211~\AA~filters reveal the highest decrease. These channels are most sensitive to temperatures around 1-2~MK, which corresponds to quiet Sun coronal temperatures. During the CME lift-off such plasma is evacuated, resulting in regions of reduced intensity (coronal dimmings). The observed intensity in SDO/AIA results from the sum of emission along the LOS. Assuming that along a specific LOS, contributions from coronal dimming regions (negative contributions) as well as the EUV wave (positive contributions) are present, the total intensity can be reduced to the background intensity, when summed up. The 335~\AA~channel measures plasma at higher temperatures around $2.5\times10^{6}$~K and is thus less sensitive to the expansion and evacuation of quiet coronal plasma by the erupting CME (cf. Fig.~\ref{fig:event}). 

2) Observations from SDO/AIA revealed signatures of different types of coronal dimmings. Using a thresholding technique we were able to distinguish potential core dimming regions from secondary/remote dimming regions. The identified core dimming regions show evacuation of dense plasma, observed as maximum decrease in intensity in the overall identified dimming region from SDO/AIA and \textbf{as the largest decrease} in emission from the ST-A LOS reconstructions. They are located in regions of opposite magnetic polarities and lie close to the eruption site (cf.~Fig.~\ref{dimming}). Therefore, we conclude that these regions point to real footpoints of the erupting fluxrope. 
For the secondary/remote dimmings (red ROI and lower part of the cyan ellipse in Fig.~\ref{dimming_los}) the reconstruction of the emission from ST-A  revealed that whether the observed intensity in SDO/AIA results in a dimming or not, depends strongly on the structures lying along the corresponding LOS. 
We conclude that projection effects play an important role in the appearance and correct interpretation of coronal dimming regions.

3) The kinematical evolution of the wave is derived to be different when measured from different vantage points. This can be explained by projection effects as shown in Fig.~\ref{fig:stereo}. Obviously, the on-disk wave signatures cover contributions from the outermost bright front of the CME eruption, when the EUV wave is not yet detached from the CME which is driving the wave. \cite{Ma:2009} and \cite{Hoilijoki:2013} found that there exist viewing angles, where the orientation of the LOS is tangential to the erupting dome, producing visible features that can be interpreted as parts of an EUV wave. 
In a recent study by \cite{Delannee:2014}, using different techniques for deriving the altitude of an EUV wave, the resulting heights span a distance range of 34--154~Mm above the solar limb, which is consistent with our result of $\sim$\,130~Mm for the peak of the wave pulse identified from SDO/AIA.

Another important aspect is that the on-disk observations of EUV waves may contain significant intensity contributions of the emission from the expanding CME body integrated along the LOS over several scale heights. The speeds of the EUV wave (as derived from SDO/AIA) and the CME (derived from ST-A with low projection effects) are similar, implying that in both instances we measure the same propagating structure.
In addition, the observed ``bending'' of the wave towards the solar East (see movie no 1.\ time range $\sim$~22:24~UT until 22:30~UT), is another feature suggestive of projection effects, as no magnetic obstacles are present at that location, that would cause a reflection or change of propagation direction. Assuming that the ``bending'' results from LOS integration along the outermost front of the CME, we derive that these contributions might come from large heights in the corona. For the time when the wave ``bending'' is clearly observed, the apex of the CME, as derived from ST-A, is at a distance of about~$2-2.5$~R$_{\odot}$.

The event studied in this paper is a good example of how the use of single-view image data may limit our ability to correctly interpret coronal features, such as EUV waves and coronal dimmings. 

\acknowledgments
We thank the referee for careful consideration of this manuscript and useful comments.
This study was funded by the Austrian Space Applications Programme of the Austrian Research Promotion Agency FFG (ASAP-11 4900217) and the Austrian Science Fund FWF (P24092-N16). K.D. is thankful to N. Nitta and K. Kozarev for helpful discussions. A.M.V. and K.V. acknowledge the fruitful discussions during the ISSI Team Meeting ``The Nature of Coronal Bright Fronts" led by David Long and Shaun Bloomfield. {\it SDO} data are courtesy of the NASA/{\it SDO} AIA and HMI science teams, {\it STEREO} data are courtesy of the NASA/{\it STEREO} SECCHI teams. The CDAW CME catalog is generated and maintained at the CDAW Data Center by NASA and The Catholic University of America in cooperation with the Naval Research Laboratory. SOHO is a project of international cooperation between ESA and NASA. This work 
utilizes data obtained by the GONG Program, managed by the National Solar Observatory, which is operated by AURA, Inc. under a cooperative agreement with the National Science Foundation. We are also grateful to the staff of the Hiraiso Radiospectrograph, Culgoora Spectrograph, Seagamore-Hill and Green Bank Solar Radio Burst Spectrometer for their open data policy.

\bibliographystyle{apj}

\begin{thebibliography}{}
\expandafter\ifx\csname natexlab\endcsname\relax\def\natexlab#1{#1}\fi

\bibitem[{{Aschwanden}(2009)}]{Aschwanden:2009}
{Aschwanden}, M.~J. 2009, Annales Geophysicae, 27, 3275

\bibitem[{{Attrill} {et~al.}(2007){Attrill}, {Harra}, {van Driel-Gesztelyi}, \&
  {D{\'e}moulin}}]{Attrill:2007}
{Attrill}, G.~D.~R., {Harra}, L.~K., {van Driel-Gesztelyi}, L., \&
  {D{\'e}moulin}, P. 2007, \apjl, 656, L101

\bibitem[{{Attrill} {et~al.}(2010){Attrill}, {Harra}, {van Driel-Gesztelyi}, \&
  {Wills-Davey}}]{Attrill:2010}
{Attrill}, G.~D.~R., {Harra}, L.~K., {van Driel-Gesztelyi}, L., \&
  {Wills-Davey}, M.~J. 2010, \solphys, 264, 119

\bibitem[{{Biesecker} {et~al.}(2002){Biesecker}, {Myers}, {Thompson}, {Hammer},
  \& {Vourlidas}}]{Biesecker:2002}
{Biesecker}, D.~A., {Myers}, D.~C., {Thompson}, B.~J., {Hammer}, D.~M., \&
  {Vourlidas}, A. 2002, \apj, 569, 1009

\bibitem[{{Bougeret} {et~al.}(1995){Bougeret}, {Kaiser}, {Kellogg}, {Manning},
  {Goetz}, {Monson}, {Monge}, {Friel}, {Meetre}, {Perche}, {Sitruk}, \&
  {Hoang}}]{Bougeret:1995}
{Bougeret}, J.-L., {Kaiser}, M.~L., {Kellogg}, P.~J., {et~al.} 1995, \ssr, 71,
  231

\bibitem[{{Bougeret} {et~al.}(2008){Bougeret}, {Goetz}, {Kaiser}, {Bale},
  {Kellogg}, {Maksimovic}, {Monge}, {Monson}, {Astier}, {Davy}, {Dekkali},
  {Hinze}, {Manning}, {Aguilar-Rodriguez}, {Bonnin}, {Briand}, {Cairns},
  {Cattell}, {Cecconi}, {Eastwood}, {Ergun}, {Fainberg}, {Hoang}, {Huttunen},
  {Krucker}, {Lecacheux}, {MacDowall}, {Macher}, {Mangeney}, {Meetre},
  {Moussas}, {Nguyen}, {Oswald}, {Pulupa}, {Reiner}, {Robinson}, {Rucker},
  {Salem}, {Santolik}, {Silvis}, {Ullrich}, {Zarka}, \&
  {Zouganelis}}]{Bougeret:2008}
{Bougeret}, J.~L., {Goetz}, K., {Kaiser}, M.~L., {et~al.} 2008, \ssr, 136, 487

\bibitem[{{Chen} {et~al.}(2002){Chen}, {Wu}, {Shibata}, \& {Fang}}]{Chen:2002}
{Chen}, P.~F., {Wu}, S.~T., {Shibata}, K., \& {Fang}, C. 2002, \apjl, 572, L99

\bibitem[{{Cheng} {et~al.}(2012){Cheng}, {Zhang}, {Olmedo}, {Vourlidas},
  {Ding}, \& {Liu}}]{Cheng:2012}
{Cheng}, X., {Zhang}, J., {Olmedo}, O., {et~al.} 2012, \apjl, 745, L5

\bibitem[{{Cho} {et~al.}(2013){Cho}, {Gopalswamy}, {Kwon}, {Kim}, \&
  {Yashiro}}]{Cho:2013}
{Cho}, K.-S., {Gopalswamy}, N., {Kwon}, R.-Y., {Kim}, R.-S., \& {Yashiro}, S.
  2013, \apj, 765, 148

\bibitem[{{Cliver} {et~al.}(2005){Cliver}, {Laurenza}, {Storini}, \&
  {Thompson}}]{Cliver:2005}
{Cliver}, E.~W., {Laurenza}, M., {Storini}, M., \& {Thompson}, B.~J. 2005,
  \apj, 631, 604

\bibitem[{{Cohen} {et~al.}(2009){Cohen}, {Attrill}, {Manchester}, \&
  {Wills-Davey}}]{Cohen:2009}
{Cohen}, O., {Attrill}, G.~D.~R., {Manchester}, IV, W.~B., \& {Wills-Davey},
  M.~J. 2009, \apj, 705, 587

\bibitem[{{Delann{\'e}e} {et~al.}(2014){Delann{\'e}e}, {Artzner}, {Schmieder},
  \& {Parenti}}]{Delannee:2014}
{Delann{\'e}e}, C., {Artzner}, G., {Schmieder}, B., \& {Parenti}, S. 2014,
  \solphys, 289, 2565

\bibitem[{{Delann{\'e}e} \& {Aulanier}(1999)}]{Delannee:1999}
{Delann{\'e}e}, C., \& {Aulanier}, G. 1999, \solphys, 190, 107

\bibitem[{{Downs} {et~al.}(2012){Downs}, {Roussev}, {van der Holst}, {Lugaz},
  \& {Sokolov}}]{Downs:2012}
{Downs}, C., {Roussev}, I.~I., {van der Holst}, B., {Lugaz}, N., \& {Sokolov},
  I.~V. 2012, \apj, 750, 134

\bibitem[{{Downs} {et~al.}(2011){Downs}, {Roussev}, {van der Holst}, {Lugaz},
  {Sokolov}, \& {Gombosi}}]{Downs:2011}
{Downs}, C., {Roussev}, I.~I., {van der Holst}, B., {et~al.} 2011, \apj, 728, 2

\bibitem[{{Feng} {et~al.}(2013){Feng}, {Wiegelmann}, {Su}, {Inhester}, {Li},
  {Sun}, \& {Gan}}]{Feng:2013}
{Feng}, L., {Wiegelmann}, T., {Su}, Y., {et~al.} 2013, \apj, 765, 37

\bibitem[{{Gallagher} \& {Long}(2011)}]{Gallagher:2011}
{Gallagher}, P.~T., \& {Long}, D.~M. 2011, \ssr, 158, 365

\bibitem[{{Harra} \& {Sterling}(2001)}]{Harra:2001}
{Harra}, L.~K., \& {Sterling}, A.~C. 2001, \apjl, 561, L215

\bibitem[{{Harrison} {et~al.}(2003){Harrison}, {Bryans}, {Simnett}, \&
  {Lyons}}]{Harrison:2003}
{Harrison}, R.~A., {Bryans}, P., {Simnett}, G.~M., \& {Lyons}, M. 2003, \aap,
  400, 1071

\bibitem[{{Harrison} \& {Lyons}(2000)}]{Harrison:2000}
{Harrison}, R.~A., \& {Lyons}, M. 2000, \aap, 358, 1097

\bibitem[{{Hoilijoki} {et~al.}(2013){Hoilijoki}, {Pomoell}, {Vainio},
  {Palmroth}, \& {Koskinen}}]{Hoilijoki:2013}
{Hoilijoki}, S., {Pomoell}, J., {Vainio}, R., {Palmroth}, M., \& {Koskinen},
  H.~E.~J. 2013, \solphys, 286, 493

\bibitem[{{Howard} {et~al.}(2008){Howard}, {Moses}, {Vourlidas}, {Newmark},
  {Socker}, {Plunkett}, {Korendyke}, {Cook}, {Hurley}, {Davila}, {Thompson},
  {St Cyr}, {Mentzell}, {Mehalick}, {Lemen}, {Wuelser}, {Duncan}, {Tarbell},
  {Wolfson}, {Moore}, {Harrison}, {Waltham}, {Lang}, {Davis}, {Eyles},
  {Mapson-Menard}, {Simnett}, {Halain}, {Defise}, {Mazy}, {Rochus}, {Mercier},
  {Ravet}, {Delmotte}, {Auchere}, {Delaboudiniere}, {Bothmer}, {Deutsch},
  {Wang}, {Rich}, {Cooper}, {Stephens}, {Maahs}, {Baugh}, {McMullin}, \&
  {Carter}}]{Howard:2008}
{Howard}, R.~A., {Moses}, J.~D., {Vourlidas}, A., {et~al.} 2008, \ssr, 136, 67

\bibitem[{{Hudson} {et~al.}(1996){Hudson}, {Acton}, \&
  {Freeland}}]{Hudson:1996}
{Hudson}, H.~S., {Acton}, L.~W., \& {Freeland}, S.~L. 1996, \apj, 470, 629

\bibitem[{{Janvier} {et~al.}(2016){Janvier}, {Savcheva}, {Pariat}, {Tassev},
  {Millholland}, {Bommier}, {McCauley}, {McKillop}, \& {Dougan}}]{Janvier:2016}
{Janvier}, M., {Savcheva}, A., {Pariat}, E., {et~al.} 2016, ArXiv e-prints,
  arXiv:1604.07241

\bibitem[{{Jiang} {et~al.}(2013){Jiang}, {Feng}, {Wu}, \& {Hu}}]{Jiang:2013}
{Jiang}, C., {Feng}, X., {Wu}, S.~T., \& {Hu}, Q. 2013, \apjl, 771, L30

\bibitem[{{Kaiser}(2005)}]{Kaiser:2005}
{Kaiser}, M.~L. 2005, Advances in Space Research, 36, 1483

\bibitem[{{Kaiser} {et~al.}(2008){Kaiser}, {Kucera}, {Davila}, {St.~Cyr},
  {Guhathakurta}, \& {Christian}}]{Kaiser:2008}
{Kaiser}, M.~L., {Kucera}, T.~A., {Davila}, J.~M., {et~al.} 2008, \ssr, 136, 5

\bibitem[{{Kienreich} {et~al.}(2009){Kienreich}, {Temmer}, \&
  {Veronig}}]{Kienreich:2009}
{Kienreich}, I.~W., {Temmer}, M., \& {Veronig}, A.~M. 2009, \apjl, 703, L118

\bibitem[{{Klassen} {et~al.}(2000){Klassen}, {Aurass}, {Mann}, \&
  {Thompson}}]{Klassen:2000}
{Klassen}, A., {Aurass}, H., {Mann}, G., \& {Thompson}, B.~J. 2000, \aaps, 141,
  357

\bibitem[{{Lemen} {et~al.}(2012){Lemen}, {Title}, {Akin}, {Boerner}, {Chou},
  {Drake}, {Duncan}, {Edwards}, {Friedlaender}, {Heyman}, {Hurlburt}, {Katz},
  {Kushner}, {Levay}, {Lindgren}, {Mathur}, {McFeaters}, {Mitchell}, {Rehse},
  {Schrijver}, {Springer}, {Stern}, {Tarbell}, {Wuelser}, {Wolfson}, {Yanari},
  {Bookbinder}, {Cheimets}, {Caldwell}, {Deluca}, {Gates}, {Golub}, {Park},
  {Podgorski}, {Bush}, {Scherrer}, {Gummin}, {Smith}, {Auker}, {Jerram},
  {Pool}, {Soufli}, {Windt}, {Beardsley}, {Clapp}, {Lang}, \&
  {Waltham}}]{Lemen:2012}
{Lemen}, J.~R., {Title}, A.~M., {Akin}, D.~J., {et~al.} 2012, \solphys, 275, 17

\bibitem[{{Liu} \& {Ofman}(2014)}]{Liu:2014}
{Liu}, W., \& {Ofman}, L. 2014, \solphys, 289, 3233

\bibitem[{{Long} {et~al.}(2008){Long}, {Gallagher}, {McAteer}, \&
  {Bloomfield}}]{Long:2008}
{Long}, D.~M., {Gallagher}, P.~T., {McAteer}, R.~T.~J., \& {Bloomfield}, D.~S.
  2008, \apjl, 680, L81

\bibitem[{{Long} {et~al.}(2011){Long}, {Gallagher}, {McAteer}, \&
  {Bloomfield}}]{Long:2011}
---. 2011, \aap, 531, A42

\bibitem[{{Ma} {et~al.}(2011){Ma}, {Raymond}, {Golub}, {Lin}, {Chen}, {Grigis},
  {Testa}, \& {Long}}]{Ma:2011}
{Ma}, S., {Raymond}, J.~C., {Golub}, L., {et~al.} 2011, \apj, 738, 160

\bibitem[{{Ma} {et~al.}(2009){Ma}, {Wills-Davey}, {Lin}, {Chen}, {Attrill},
  {Chen}, {Zhao}, {Li}, \& {Golub}}]{Ma:2009}
{Ma}, S., {Wills-Davey}, M.~J., {Lin}, J., {et~al.} 2009, \apj, 707, 503

\bibitem[{{Mandrini} {et~al.}(2007){Mandrini}, {Nakwacki}, {Attrill}, {van
  Driel-Gesztelyi}, {D{\'e}moulin}, {Dasso}, \& {Elliott}}]{Mandrini:2007}
{Mandrini}, C.~H., {Nakwacki}, M.~S., {Attrill}, G., {et~al.} 2007, \solphys,
  244, 25

\bibitem[{{Mandrini} {et~al.}(2005){Mandrini}, {Pohjolainen}, {Dasso}, {Green},
  {D{\'e}moulin}, {van Driel-Gesztelyi}, {Copperwheat}, \&
  {Foley}}]{Mandrini:2005}
{Mandrini}, C.~H., {Pohjolainen}, S., {Dasso}, S., {et~al.} 2005, \aap, 434,
  725

\bibitem[{{Markwardt}(2009)}]{Markwardt:2009}
{Markwardt}, C.~B. 2009, in Astronomical Society of the Pacific Conference
  Series, Vol. 411, Astronomical Data Analysis Software and Systems XVIII, ed.
  D.~A. {Bohlender}, D.~{Durand}, \& P.~{Dowler}, 251

\bibitem[{{Miklenic} {et~al.}(2011){Miklenic}, {Veronig}, {Temmer},
  {M{\"o}stl}, \& {Biernat}}]{Miklenic:2011}
{Miklenic}, C., {Veronig}, A.~M., {Temmer}, M., {M{\"o}stl}, C., \& {Biernat},
  H.~K. 2011, \solphys, 273, 125

\bibitem[{{Moreton}(1960)}]{Moreton:1960}
{Moreton}, G.~E. 1960, \aj, 65, 494

\bibitem[{{Muhr} {et~al.}(2011){Muhr}, {Veronig}, {Kienreich}, {Temmer}, \&
  {Vr{\v s}nak}}]{Muhr:2011}
{Muhr}, N., {Veronig}, A.~M., {Kienreich}, I.~W., {Temmer}, M., \& {Vr{\v
  s}nak}, B. 2011, \apj, 739, 89

\bibitem[{{Muhr} {et~al.}(2014){Muhr}, {Veronig}, {Kienreich}, {Vr{\v s}nak},
  {Temmer}, \& {Bein}}]{Muhr:2014}
{Muhr}, N., {Veronig}, A.~M., {Kienreich}, I.~W., {et~al.} 2014, \solphys, 289,
  4563

\bibitem[{{Nitta} {et~al.}(2014){Nitta}, {Liu}, {Gopalswamy}, \&
  {Yashiro}}]{Nitta:2014}
{Nitta}, N.~V., {Liu}, W., {Gopalswamy}, N., \& {Yashiro}, S. 2014, \solphys,
  289, 4589

\bibitem[{{Nitta} {et~al.}(2013){Nitta}, {Schrijver}, {Title}, \&
  {Liu}}]{Nitta:2013}
{Nitta}, N.~V., {Schrijver}, C.~J., {Title}, A.~M., \& {Liu}, W. 2013, \apj,
  776, 58

\bibitem[{{Ofman} \& {Thompson}(2002)}]{Ofman:2002}
{Ofman}, L., \& {Thompson}, B.~J. 2002, \apj, 574, 440

\bibitem[{{Olmedo} {et~al.}(2012){Olmedo}, {Vourlidas}, {Zhang}, \&
  {Cheng}}]{Olmedo:2012}
{Olmedo}, O., {Vourlidas}, A., {Zhang}, J., \& {Cheng}, X. 2012, \apj, 756, 143

\bibitem[{{Patsourakos} \& {Vourlidas}(2009)}]{Patsourakos:2009a}
{Patsourakos}, S., \& {Vourlidas}, A. 2009, \apjl, 700, L182

\bibitem[{{Patsourakos} \& {Vourlidas}(2012)}]{Patsourakos:2012}
---. 2012, \solphys, 281, 187

\bibitem[{{Patsourakos} {et~al.}(2009){Patsourakos}, {Vourlidas}, {Wang},
  {Stenborg}, \& {Thernisien}}]{Patsourakos:2009}
{Patsourakos}, S., {Vourlidas}, A., {Wang}, Y.~M., {Stenborg}, G., \&
  {Thernisien}, A. 2009, \solphys, 259, 49

\bibitem[{{Pesnell} {et~al.}(2012){Pesnell}, {Thompson}, \&
  {Chamberlin}}]{Pesnell:2012}
{Pesnell}, W.~D., {Thompson}, B.~J., \& {Chamberlin}, P.~C. 2012, \solphys,
  275, 3

\bibitem[{{Podladchikova} \& {Berghmans}(2005)}]{Podladchikova:2005}
{Podladchikova}, O., \& {Berghmans}, D. 2005, \solphys, 228, 265

\bibitem[{{Pomoell} {et~al.}(2008){Pomoell}, {Vainio}, \&
  {Kissmann}}]{Pomoell:2008}
{Pomoell}, J., {Vainio}, R., \& {Kissmann}, R. 2008, \solphys, 253, 249

\bibitem[{{Romano} {et~al.}(2015){Romano}, {Zuccarello}, {Guglielmino},
  {Berrilli}, {Bruno}, {Carbone}, {Consolini}, {de Lauretis}, {Del Moro},
  {Elmhamdi}, {Ermolli}, {Fineschi}, {Francia}, {Kordi}, {Landi
  Degl'Innocenti}, {Laurenza}, {Lepreti}, {Marcucci}, {Pallocchia},
  {Pietropaolo}, {Romoli}, {Vecchio}, {Vellante}, \& {Villante}}]{Romano:2015}
{Romano}, P., {Zuccarello}, F., {Guglielmino}, S.~L., {et~al.} 2015, \aap, 582,
  A55

\bibitem[{{Saito} {et~al.}(1970){Saito}, {Makita}, {Nishi}, \&
  {Hata}}]{Saito:1970}
{Saito}, K., {Makita}, M., {Nishi}, K., \& {Hata}, S. 1970, Annals of the Tokyo
  Astronomical Observatory, 12, 53

\bibitem[{{Schrijver} {et~al.}(2011){Schrijver}, {Aulanier}, {Title}, {Pariat},
  \& {Delann{\'e}e}}]{Schrijver:2011}
{Schrijver}, C.~J., {Aulanier}, G., {Title}, A.~M., {Pariat}, E., \&
  {Delann{\'e}e}, C. 2011, \apj, 738, 167

\bibitem[{{Shen} {et~al.}(2014){Shen}, {Ichimoto}, {Ishii}, {Tian}, {Zhao}, \&
  {Shibata}}]{Shen:2014}
{Shen}, Y., {Ichimoto}, K., {Ishii}, T.~T., {et~al.} 2014, \apj, 786, 151

\bibitem[{{Sterling} \& {Hudson}(1997)}]{Sterling:1997}
{Sterling}, A.~C., \& {Hudson}, H.~S. 1997, \apjl, 491, L55

\bibitem[{{Temmer} {et~al.}(2011){Temmer}, {Veronig}, {Gopalswamy}, \&
  {Yashiro}}]{Temmer:2011}
{Temmer}, M., {Veronig}, A.~M., {Gopalswamy}, N., \& {Yashiro}, S. 2011,
  \solphys, 273, 421

\bibitem[{{Temmer} {et~al.}(2010){Temmer}, {Veronig}, {Kontar}, {Krucker}, \&
  {Vr{\v s}nak}}]{Temmer:2010}
{Temmer}, M., {Veronig}, A.~M., {Kontar}, E.~P., {Krucker}, S., \& {Vr{\v
  s}nak}, B. 2010, \apj, 712, 1410

\bibitem[{{Thompson} {et~al.}(2000){Thompson}, {Cliver}, {Nitta},
  {Delann{\'e}e}, \& {Delaboudini{\`e}re}}]{Thompson:2000}
{Thompson}, B.~J., {Cliver}, E.~W., {Nitta}, N., {Delann{\'e}e}, C., \&
  {Delaboudini{\`e}re}, J.-P. 2000, \grl, 27, 1431

\bibitem[{{Thompson} \& {Myers}(2009)}]{Thompson:2009}
{Thompson}, B.~J., \& {Myers}, D.~C. 2009, \apjs, 183, 225

\bibitem[{{Thompson} {et~al.}(1998){Thompson}, {Plunkett}, {Gurman}, {Newmark},
  {St.~Cyr}, \& {Michels}}]{Thompson:1998}
{Thompson}, B.~J., {Plunkett}, S.~P., {Gurman}, J.~B., {et~al.} 1998, \grl, 25,
  2465

\bibitem[{{Thompson} {et~al.}(1999){Thompson}, {Gurman}, {Neupert}, {Newmark},
  {Delaboudini{\`e}re}, {Cyr}, {Stezelberger}, {Dere}, {Howard}, \&
  {Michels}}]{Thompson:1999}
{Thompson}, B.~J., {Gurman}, J.~B., {Neupert}, W.~M., {et~al.} 1999, \apjl,
  517, L151

\bibitem[{{Thompson}(2009)}]{Thompson:2009b}
{Thompson}, W.~T. 2009, \icarus, 200, 351

\bibitem[{{Thompson} \& {Wei}(2010)}]{Thompson:2010}
{Thompson}, W.~T., \& {Wei}, K. 2010, \solphys, 261, 215

\bibitem[{{Tian} {et~al.}(2012){Tian}, {McIntosh}, {Xia}, {He}, \&
  {Wang}}]{Tian:2012}
{Tian}, H., {McIntosh}, S.~W., {Xia}, L., {He}, J., \& {Wang}, X. 2012, \apj,
  748, 106

\bibitem[{{Vanninathan} {et~al.}(2015){Vanninathan}, {Veronig}, {Dissauer},
  {Madjarska}, {Hannah}, \& {Kontar}}]{Vanninathan:2015}
{Vanninathan}, K., {Veronig}, A.~M., {Dissauer}, K., {et~al.} 2015, \apj, 812,
  173

\bibitem[{{Veronig} {et~al.}(2011){Veronig}, {G{\"o}m{\"o}ry}, {Kienreich},
  {Muhr}, {Vr{\v s}nak}, {Temmer}, \& {Warren}}]{Veronig:2011}
{Veronig}, A.~M., {G{\"o}m{\"o}ry}, P., {Kienreich}, I.~W., {et~al.} 2011,
  \apjl, 743, L10

\bibitem[{{Veronig} {et~al.}(2010){Veronig}, {Muhr}, {Kienreich}, {Temmer}, \&
  {Vr{\v s}nak}}]{Veronig:2010}
{Veronig}, A.~M., {Muhr}, N., {Kienreich}, I.~W., {Temmer}, M., \& {Vr{\v
  s}nak}, B. 2010, \apjl, 716, L57

\bibitem[{{Veronig} {et~al.}(2008){Veronig}, {Temmer}, \& {Vr{\v
  s}nak}}]{Veronig:2008}
{Veronig}, A.~M., {Temmer}, M., \& {Vr{\v s}nak}, B. 2008, \apjl, 681, L113

\bibitem[{{Wang} {et~al.}(2002){Wang}, {Yan}, {Wang}, {Kurokawa}, \&
  {Shibata}}]{Wang:2002}
{Wang}, T., {Yan}, Y., {Wang}, J., {Kurokawa}, H., \& {Shibata}, K. 2002, \apj,
  572, 580

\bibitem[{{Wang}(2000)}]{Wang:2000}
{Wang}, Y.-M. 2000, \apjl, 543, L89

\bibitem[{{Warmuth}(2010)}]{Warmuth:2010}
{Warmuth}, A. 2010, Advances in Space Research, 45, 527

\bibitem[{Warmuth(2015)}]{Warmuth:2015}
Warmuth, A. 2015, Living Reviews in Solar Physics, 12, doi:10.12942/lrsp-2015-3

\bibitem[{{Warmuth} {et~al.}(2001){Warmuth}, {Vr{\v s}nak}, {Aurass}, \&
  {Hanslmeier}}]{Warmuth:2001}
{Warmuth}, A., {Vr{\v s}nak}, B., {Aurass}, H., \& {Hanslmeier}, A. 2001,
  \apjl, 560, L105

\bibitem[{{Warmuth} {et~al.}(2004){Warmuth}, {Vr{\v s}nak}, {Magdaleni{\'c}},
  {Hanslmeier}, \& {Otruba}}]{Warmuth:2004}
{Warmuth}, A., {Vr{\v s}nak}, B., {Magdaleni{\'c}}, J., {Hanslmeier}, A., \&
  {Otruba}, W. 2004, \aap, 418, 1117

\bibitem[{{Webb} {et~al.}(2000){Webb}, {Lepping}, {Burlaga}, {DeForest},
  {Larson}, {Martin}, {Plunkett}, \& {Rust}}]{Webb:2000}
{Webb}, D.~F., {Lepping}, R.~P., {Burlaga}, L.~F., {et~al.} 2000, \jgr, 105,
  27251

\bibitem[{{Wills-Davey} \& {Thompson}(1999)}]{Wills-Davey:1999}
{Wills-Davey}, M.~J., \& {Thompson}, B.~J. 1999, \solphys, 190, 467

\bibitem[{{Wu} {et~al.}(2001){Wu}, {Zheng}, {Wang}, {Thompson}, {Plunkett},
  {Zhao}, \& {Dryer}}]{Wu:2001}
{Wu}, S.~T., {Zheng}, H., {Wang}, S., {et~al.} 2001, \jgr, 106, 25089

\bibitem[{{Zarro} {et~al.}(1999){Zarro}, {Sterling}, {Thompson}, {Hudson}, \&
  {Nitta}}]{Zarro:1999}
{Zarro}, D.~M., {Sterling}, A.~C., {Thompson}, B.~J., {Hudson}, H.~S., \&
  {Nitta}, N. 1999, \apjl, 520, L139

\bibitem[{{Zhukov}(2011)}]{Zhukov:2011}
{Zhukov}, A.~N. 2011, Journal of Atmospheric and Solar-Terrestrial Physics, 73,
  1096

\bibitem[{{Zhukov} \& {Auch{\`e}re}(2004)}]{Zhukov:2004}
{Zhukov}, A.~N., \& {Auch{\`e}re}, F. 2004, \aap, 427, 705

\end{thebibliography}

\end{document}